%% file: main.tex
\documentclass[journal]{IEEEtran}

\usepackage{times}
\usepackage{epsfig}
\usepackage{graphicx}
\usepackage{amsmath}
\usepackage{amssymb}
\usepackage{colortbl}
\usepackage{multirow}
\usepackage{soul}
\usepackage{adjustbox}
\usepackage{array}
\usepackage{setspace}
\usepackage{lscape}
\usepackage{mathtools}
\usepackage{dblfloatfix} % figures at bottom page
\usepackage{hyperref}
\usepackage{pgfplots}
\pgfplotsset{compat=newest}

\usepackage{enumitem}
\usepackage{multirow}
\usepackage{rotating}
\usepackage{caption}
\usepackage{cite}
\usepackage{amsmath,amssymb,amsfonts}
\usepackage{algorithmic}
\usepackage{graphicx}
\usepackage{textcomp}
\usepackage{caption} % DO NOT CHANGE THIS AND DO NOT ADD ANY OPTIONS TO IT
\usepackage{subcaption}
\usepackage{amssymb}% http://ctan.org/pkg/amssymb
\usepackage{pifont}% http://ctan.org/pkg/pifont
\usepackage{amsmath,bm}

\usepackage{fixfoot,xspace}

\DeclareFixedFootnote*{\dbfootnote}{\url{https://github.com/BiDAlab/ChildCIdb_v1}}

\begin{document}

\title{Children Age Group Detection based on Human-Computer Interaction and \\ Time Series Analysis}

\author{
Juan Carlos Ruiz-Garcia, Carlos Hojas, Ruben Tolosana, Ruben Vera-Rodriguez, \\ Aythami Morales, Julian Fierrez, Javier Ortega-Garcia, Jaime Herreros-Rodriguez\\

\thanks{J.C. Ruiz-Garcia, Carlos Hojas, R. Tolosana, R. Vera-Rodriguez, A. Morales and J. Fierrez are with the Biometrics and Data Pattern Analytics - BiDA Lab, Escuela Politecnica Superior, Universidad Autonoma de Madrid, 28049 Madrid, Spain (e-mail: juanc.ruiz@uam.es, carlos.hojas@estudiante.uam.es; ruben.tolosana@uam.es; ruben.vera@uam.es; aythami.morales@uam.es; julian.fierrez@uam.es; javier.ortega@uam.es).

J. Herreros-Rodriguez is with the Hospital Universitario Infanta Leonor, 28031 Madrid, Spain (e-mail: hrinvest@hotmail.com).}}% <-this % stops a space

\maketitle

% ABSTRACT SECTION
\input{body/1_abstract}

% KEYWORDS SECTION
\input{body/2_keywords}

% INTRODUCTION SECTION
\input{body/3_introduction}

% RELATED WORKS SECTION
\input{body/4_related_works}

% DATABASE SECTION
\input{body/5_childcidb}

% PROPOSED METHODS SECTION
\input{body/6_proposed_method}

% EXPERIMENTS AND RESULTS SECTION
\input{body/7_experiments_results}
 
% CONCLUSION AND FUTURE WORKS SECTION
\input{body/8_conclusion_feature_works}

% ACKNOWLEDGEMENTS SECTION
\input{body/9_acknowledgements}

% REFERENCES SECTION

{
\bibliographystyle{IEEEtran}
\bibliography{main.bib}
}

% BIOGRAPHY SECTION
\input{body/10_biography}

\end{document}

%% file: body/1_abstract.tex
\begin{abstract}
    This article proposes a novel Children-Computer Interaction (CCI) approach for the task of age group detection. This approach focuses on the automatic analysis of the time series generated from the interaction of the children with mobile devices. In particular, we extract a set of 25 time series related to spatial, pressure, and kinematic information of the children interaction while colouring a tree through a pen stylus tablet, a specific test from the large-scale public ChildCIdb database\dbfootnote.

    A complete analysis of the proposed approach is carried out using different time series selection techniques to choose the most discriminative ones for the age group detection task: \textit{i)} a statistical analysis, and \textit{ii)} an automatic algorithm called Sequential Forward Search (SFS). In addition, different classification algorithms such as Dynamic Time Warping Barycenter Averaging (DBA) and Hidden Markov Models (HMM) are studied. Accuracy results over 85\% are achieved, outperforming previous approaches in the literature and in more challenging age group conditions. Finally, the approach presented in this study can benefit many children-related applications, for example, towards an age-appropriate environment with the technology.
\end{abstract}

%% file: body/2_keywords.tex
% Note that keywords are not normally used for peer-review papers.

%% file: body/3_introduction.tex
\section{Introduction}
\IEEEPARstart{C}{hildren}'s exposure to mobile devices has increased dramatically in recent decades due to technological innovation~\cite{Radesky2020}. They are growing up in environments overloaded with multiple digital technologies such as smartphones, tablets, and smart TVs, among others. In addition, parents often let their children use mobile devices to keep them calm in public places, after school homework or household chores, and at bedtime before going to sleep~\cite{Kabali2015}.  As a result, the average daily usage of mobile devices in children aged 0-8 years has increased over 11 times from 2011 to 2020~\cite{Rideout2020}.

Recent studies in the literature highlight how the correct use of mobile devices can positively affect children's development and learning (e.g., through educational games or creative applications~\cite{Lawrence2021}). For example, Huber \textit{et al.} conducted a study in~\cite{Huber2016} in which children learned to solve the popular game ``Tower of Hanoi'' using a touchscreen device and subsequently apply this learning with a physical replica of the game. The results concluded that, for certain activities, children are quite capable of transferring learning from touchscreen devices to real-life problems. Language enhancement through mobile interaction games was also studied in~\cite{Dore2019}. In that work, children aged 4 years old played with an interactive word-learning application on a mobile device. The experiments concluded that children who played with the application gained a receptive and expressive understanding of the target words. With respect to self-regulation development, an interesting study was presented in~\cite{Huber2018}. The authors reported that when children from 2 to 3 years old played with an educational app for an appropriate amount of time, their self-regulation scores were higher than, for example, after exposure to watching cartoons on TV.

\begin{figure*}[!]
    \begin{center}
       \includegraphics[width=\linewidth]{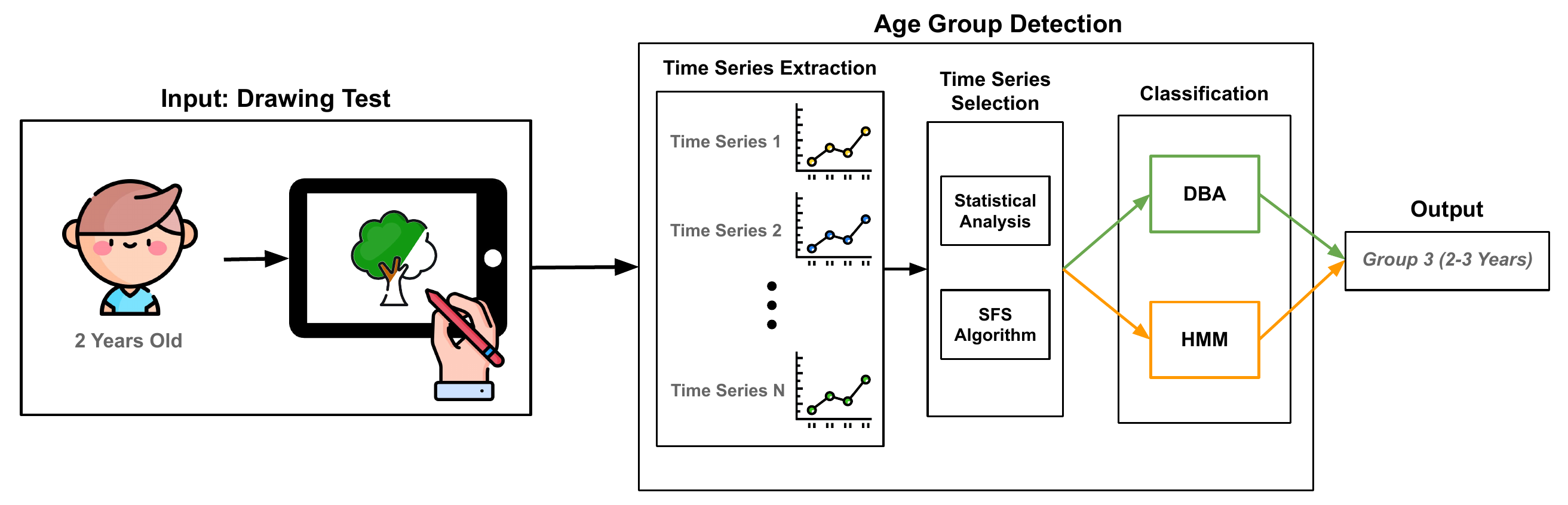}
    \end{center}
    \caption{Architecture of the proposed approach to detect the children's age group through the interaction of the children with mobile devices and the automatic analysis of the time series generated. Blue dashed arrows represent the different configurations studied in this article in terms of Time Series and Classification. First, children perform the input stage by colouring a tree on a tablet device using a stylus as an acquisition tool (Drawing Test). Next, time series extraction and selection are performed. Finally, to predict the children's age group, two different classifiers are tested independently, DBA and HMM.}
    \label{fig:architecture}
\end{figure*}

However, intensive and unsupervised exposure to digital devices in childhood can also be associated with adverse effects on growth (e.g., learning, self-regulation, well-being, social-emotional skills, sleep, media addiction, etc.). For example, Bozzola \textit{et al.} stated in~\cite{Bozzola2018} that excessive touchscreen use can affect children's correct development due to the lack of real experiences that challenge their thinking and problem-solving skills. Another interesting work in this line was presented in~\cite{Chiang2016}. The authors analysed the correlation between tablet use and physical discomforts, such as headaches and neck and shoulder pain. The results revealed that poor posture while using tablets significantly reduces the flexion angles of the head and neck. As a result, correct posture is essential for children's well-being. The associations between children's socio-emotional development and mobile media use were explored in~\cite{Radesky2020}. Results showed that children with social-emotional difficulties are more likely to be given mobile devices as a calming tool. Similar conclusions were obtained in~\cite{Radesky2014}, stating that one of the reasons for children having self-regulation difficulties (e.g., self-control, sleep, emotional regulation, and attention) is related to high media exposure at 2 years old. The effect of children mobile interaction on the sleep quality was also studied in~\cite{Cespedes2014}. In that work, the authors showed how increased exposure to media devices in the bedroom during childhood can reduce sleep duration by forming long-lasting habits that lead to significant sleep deficits. Finally, Csibi \textit{et al.} analysed in~\cite{Csibi2021} the risk of smartphone addiction in different age groups and concluded that children and young adults are at the highest risk of addictive behavior. Smartphone addiction can negative affect children's mental health, well-being, and academic performance~\cite{Samaha2016}.

All in all, the interaction of children with mobile devices can have both positive and negative effects, depending on the scenario considered. As a result, it seems critical to provide an age-appropriate environment for children (e.g., limiting exposure time to devices depending on the age, assisting children with specific touch input, preventing access to age-inappropriate applications or websites, etc.) in order to improve their development and creative skills, among others~\cite{Huber2016, Lawrence2021}. This is specially important as more and more children are exposed to mobile devices for longer time and at a younger age.

% Tabla con trabajos relacionados
\begin{table*}[t]
\caption{Comparison of studies focused on the age detection of children through their interaction with mobile devices. AGD performance refers to the \textit{Age Group Distance} presented in this study (see Sec.~\ref{classification_metrics}).}
\label{tab:comparison}
\resizebox{\linewidth}{!}{%
    \begin{tabular}{ccccccc}
    \textbf{Study} & \textbf{Detection Task} & \textbf{Age of Children} & \textbf{\begin{tabular}[c]{@{}c@{}}\# Participants\\ (Children / Adults)\end{tabular}} & \textbf{Acquisition Tool} & \textbf{Approach} & \textbf{Performance} \\ \hline
    \begin{tabular}[c]{@{}c@{}}Kim \textit{et al.} (2013)\\ \cite{Kim2013}\end{tabular}\rule{0pt}{20pt} & Child-Adult & 3-8 Years & \begin{tabular}[c]{@{}c@{}}24\\ (20/4)\end{tabular} & Stylus & Global Features & Accuracy = 90.4\% \\
    \begin{tabular}[c]{@{}c@{}}Vatavu \textit{et al.} (2015)\\ \cite{Vatavu2015}\end{tabular}\rule{0pt}{17pt} & Child-Adult & 3-6 Years & \begin{tabular}[c]{@{}c@{}}119 \\ (89/30)\end{tabular} & Finger & Global Features & Accuracy = 86.5\% \\
    \begin{tabular}[c]{@{}c@{}}Davarci \textit{et al.} (2017)\\ \cite{Davarci2017}\end{tabular}\rule{0pt}{17pt} & Child-Adult & 3-11 Years & \begin{tabular}[c]{@{}c@{}}200\\ (100/100)\end{tabular} & \begin{tabular}[c]{@{}c@{}}Finger\\ Built-in Sensors\end{tabular} & Global Features & Accuracy = 92.5\% \\
    \begin{tabular}[c]{@{}c@{}}Li \textit{et al.} (2018)\\ \cite{Li2018}\end{tabular}\rule{0pt}{17pt} & Child-Adult & 3-11 Years & \begin{tabular}[c]{@{}c@{}}31\\ (17/14)\end{tabular} & Finger & Global Features & Accuracy = 97\% \\
    \begin{tabular}[c]{@{}c@{}}Nguyen \textit{et al.} (2019)\\ \cite{Nguyen2019}\end{tabular}\rule{0pt}{17pt} & Child-Adult & 3-12 Years & \begin{tabular}[c]{@{}c@{}}50\\ (25/25)\end{tabular} & \begin{tabular}[c]{@{}c@{}}Finger\\ Built-in Sensors\end{tabular} & Global Features & Accuracy = 99\% \\
    \begin{tabular}[c]{@{}c@{}}Acien \textit{et al.} (2019)\\ \cite{Acien2019}\end{tabular}\rule{0pt}{17pt} & Child-Adult & 3-6 Years & \begin{tabular}[c]{@{}c@{}}119\\ (89/30)\end{tabular} & Finger & Global Features & Accuracy = 96.5\% \\
    \begin{tabular}[c]{@{}c@{}}Vera-Rodriguez \textit{et al.} (2020)\\ \cite{VeraRodriguez2020}\end{tabular}\rule{0pt}{17pt} & Child-Adult & 3-6 Years & \begin{tabular}[c]{@{}c@{}}119\\ (89/30)\end{tabular} & Finger & Global Features & Accuracy = 96.3\% \\
    \begin{tabular}[c]{@{}c@{}}Tolosana \textit{et al.} (2022)\\ \cite{Tolosana2022a}\end{tabular}\rule{0pt}{18pt} & \begin{tabular}[c]{@{}c@{}}Children Age Group\\ (3 Groups)\end{tabular} & 18 Months - 8 Years & \begin{tabular}[c]{@{}c@{}}438\\ (438/0)\end{tabular} & Stylus & Global Features & \begin{tabular}[c]{@{}c@{}}Accuracy = 90.45\%\\ (Drawing Test)\end{tabular} \\
    \begin{tabular}[c]{@{}c@{}}Ruiz-Garcia \textit{et al.} (2022)\\ \cite{Ruiz-Garcia2022}\end{tabular}\rule{0pt}{20pt} & \begin{tabular}[c]{@{}c@{}}Children Age Group\\ (3 Groups)\end{tabular} & 18 Months - 8 Years & \begin{tabular}[c]{@{}c@{}}438\\ (438/0)\end{tabular} & \begin{tabular}[c]{@{}c@{}}Finger\\ Stylus\end{tabular} & Global Features & \begin{tabular}[c]{@{}c@{}}Accuracy = 93.65\%\\ (3-Test Combination)\end{tabular} \\[+10pt] \hline
    \textbf{Present Study}\rule{0pt}{25pt} & \textbf{\begin{tabular}[c]{@{}c@{}}Children Age Group\\ (7 Groups)\end{tabular}} & \textbf{18 Months - 8 Years} & \textbf{\begin{tabular}[c]{@{}c@{}}492\\ (492/0)\end{tabular}} & \textbf{Stylus} & \textbf{Time Series} & \textbf{\begin{tabular}[c]{@{}c@{}}Accuracy = 85.39\%\\ Average AGD = 0.173\\ (Drawing Test)\end{tabular}} \\[+15pt] \hline
    \end{tabular}
}
\end{table*}

% Tabla con relación entre edad de los niños y nivel educacional
\begin{table*}[t]
\caption{Relationship between children's age range and their educational level group according to the Spanish system.}
\label{tab:age_range_group}
\resizebox{\linewidth}{!}{%
    \begin{tabular}{c|c|c|c|c|c|c|c}
    \textbf{Age Range} & 18 Months - 2 Years & 2-3 Years & 3-4 Years & 4-5 Years & 5-6 Years & 6-7 Years & 7-8 Years \\[+4pt] \hline
    \textbf{Educational Level}\rule{0pt}{12pt} & Group 2 & Group 3 & Group 4 & Group 5 & Group 6 & Group 7 & Group 8 \\
    \end{tabular}
}
\end{table*}

This study aims to advance in this research line by presenting a novel approach for the task of children age group detection based on the automatic analysis of time series (commonly known as local features~\cite{Martinez2014}) generated from the interaction of the children with mobile devices. 

The main contributions of the present work are:

\begin{itemize}
    \item An in-depth analysis of state-of-the-art approaches for the task of children age group detection through their interaction with mobile devices, remarking key public databases and results.
    \item Proposal of a novel Children-Computer Interaction (CCI) approach for the task of age group detection. This approach focuses on the automatic analysis of the time series generated from the interaction of the children with mobile devices. In particular, our proposed approach extracts a set of 25 time series related to spatial, pressure, and kinematic information of the children interaction while colouring a tree through a pen stylus on a tablet device. Fig.~\ref{fig:architecture} shows the graphical architecture of the proposed approach.
    \item A complete analysis of the proposed approach is carried out using two different feature selection techniques to choose the most discriminative ones for the age group detection task: \textit{i)} a statistical analysis, and \textit{ii)} an automatic algorithm called Sequential Forward Search (SFS). In addition, different classification algorithms such as DTW Barycenter Averaging (DBA) and Hidden Markov Models (HMM) are studied.
    \item Our proposed approach achieves accuracy results over 85\%, outperforming previous approaches in the literature and in more challenging age group conditions. Finally, the approach presented in this study can benefit many children-related applications, for example, towards an age-appropriate environment with the technology. 
\end{itemize}

The remainder of the article is organised as follows. Sec.~\ref{related_works} presents an overview of recent works studying different approaches for the task of children age group detection through mobile computer interaction. Sec.~\ref{childcidb} describes the database used in the experimental work of this study. In Sec.~\ref{proposed_method}, we present the time series and selection techniques, and the classification algorithms considered. Sec.~\ref{experiments} describes the experimental protocol and results obtained for the task of children age group detection, as well as a comparison with the state of the art. Finally, Sec.~\ref{conclusions} presents the conclusions and future research lines.

%% file: body/4_related_works.tex
\section{Related Works}\label{related_works}

In the existing literature, different studies have evaluated the age detection of children through their interaction with mobile devices. Table~\ref{tab:comparison} shows a comparison of the most relevant studies ordered by date. Most studies have focused on the detection of two age groups: children from adults.  Vatavu \textit{et al.} released in~\cite{Vatavu2015} a dataset comprising smartphone touch interaction data from 30 adults and 89 children aged 3-6 years old. In that work, the authors achieved 86.5\% accuracy in detecting children from adults using a set of features extracted from touch spatial coordinates \textit{x} and \textit{y}. Vera-Rodriguez \textit{et al.} also considered this database in their experimental protocol~\cite{VeraRodriguez2020}. They presented a set of neuromotor-skill features to detect children from adults. Results showed the discriminative ability of the proposed system with correct classification rates over 96\%. An interesting article in this line was presented by Li \textit{et al.}~\cite{Li2018}. The study aimed to investigate the different screen-touch patterns between children and adult users. Experiments were carried out using 17 children (3 to 11 years) and 14 adults (22 to 60 years), achieving 97\% accuracy for child-adult detection using only a single finger and 8 consecutive swipes on the touch screen.

% Tabla con el número de niños capturados en cada adquisición
\begin{table*}[t]
\caption{Description of the number of children who participated in each group and acquisition in the ChildCIdb database. There is no age overlap between the different groups.}
\label{tab:acquisitions}
\begin{tabular}{c|ccccccc|c}
 & \textbf{\begin{tabular}[c]{@{}c@{}}Group 2\\ (18M-2Y)\end{tabular}} & \textbf{\begin{tabular}[c]{@{}c@{}}Group 3\\ (2Y-3Y)\end{tabular}} & \textbf{\begin{tabular}[c]{@{}c@{}}Group 4\\ (3Y-4Y)\end{tabular}} & \textbf{\begin{tabular}[c]{@{}c@{}}Group 5\\ (4Y-5Y)\end{tabular}} & \textbf{\begin{tabular}[c]{@{}c@{}}Group 6\\ (5Y-6Y)\end{tabular}} & \textbf{\begin{tabular}[c]{@{}c@{}}Group 7\\ (6Y-7Y)\end{tabular}} & \textbf{\begin{tabular}[c]{@{}c@{}}Group 8\\ (7Y-8Y)\end{tabular}} & \textbf{\begin{tabular}[c]{@{}c@{}}\# Children by\\ Acquisition\end{tabular}} \\ \hline
\textbf{\begin{tabular}[c]{@{}c@{}}1st Acquisition\\ (Jan 2020)\end{tabular}} & 18 & 36 & 50 & 66 & 93 & 77 & 98 & 438 \\
\textbf{\begin{tabular}[c]{@{}c@{}}2nd Acquisition\\ (May 2021)\end{tabular}} & 40 & 18 & 36 & 51 & 67 & 89 & 75 & 376 \\
\textbf{\begin{tabular}[c]{@{}c@{}}3rd Acquisition\\ (Oct 2021)\end{tabular}} & 14 & 45 & 18 & 34 & 49 & 67 & 88 & 315 \\ \hline
\textbf{\begin{tabular}[c]{@{}c@{}}\# Children by\\ Group\end{tabular}}\rule{0pt}{15pt} & 72 & 99 & 104 & 151 & 209 & 233 & 261 &
\end{tabular}
\end{table*}

A different approach was presented by Nguyen \textit{et al.} in~\cite{Nguyen2019}. In that work, the authors tried to predict if a child or an adult is using a smartphone based on behavioral differences extracted from the touchscreen and built-in sensors (accelerometer and gyroscope). An accuracy over 99\% was achieved after 5 seconds of sensor reading or 8 consecutive touch gestures. Similar conclusions were obtained by Davarci \textit{et al.} in~\cite{Davarci2017}, where rates over 92\% were achieved in detecting children from adults based only on accelerometer data.

The user interaction with mobile devices is not only studied with the finger as input but also using a stylus~\cite{Tolosana2021, Tolosana2022b}. In line with the task of children and adults detection, Kim \textit{et al.} proposed in~\cite{Kim2013} a sketch recognition system able to assess children's developmental skills from their sketches created with a stylus on a tablet. In total, 20 children (3 to 8 years) and 4 adults participated in the study. Two main significant differences between children and adults were found: \textit{i)} the stroke lengths of adults' sketches were larger, and \textit{ii)} adults took less time to draw sketches.

Focusing on the stylus scenarios, only a few recent studies go one step further in the task of children age group detection, trying to detect the children's age group rather than just differentiating them from adults. Tolosana \textit{et al.} released in~~\cite{Tolosana2022a} a public database called ChildCIdb, including over 400 children aged 18 months to 8 years interacting with mobile devices (stylus and finger). The authors grouped the children into 3 age groups without overlap: 1 to 3 years, 3 to 6 years, and 6 to 8 years. Accurate results were obtained between different children age groups (over 90\% accuracy) by analysing only one task in which children had to colour a tree using a stylus. It is important to highlight that the children age group approach was based on a set of 148 global features, not time series analysis as in this study. Ruiz-Garcia \textit{et al.} obtained similar conclusions in~\cite{Ruiz-Garcia2022}, where authors also considered ChildCIdb as an experimental database. They analysed the correlation between children's chronological age and their motor and cognitive development while interacting with all the tests included in ChildCIdb. Children were grouped into the same three age groups considered in~\cite{Tolosana2022a}. Good results above 93\% accuracy were achieved in the age group detection task by combining three different ChildCIdb tests (i.e., Drag and Drop, Spiral Test, and Drawing Test), using both stylus and finger as input.

In the present article, we propose a novel children age group detection based on the automatic analysis of time series, different compared with previous approaches in the literature (based on global features). In addition, and as we have seen along this section, most studies in the literature focus on detecting children from adults or grouping children in wide age groups. We go deeper by performing a more complex task: detecting the age group of the children based on their educational level (7 different groups in ChildCIdb~\cite{Tolosana2015a}), i.e., one year precision. Table~\ref{tab:age_range_group} shows the relationship between children's age range and their educational level groups based on the Spanish education system.

%% file: body/5_childcidb.tex
\section{Database: ChildCIdb}\label{childcidb}

The experimental work of this study considers the public ChildCIdb database. Tolosana \textit{et al.} presented in~\cite{Tolosana2022a} the first version of this database (ChildCIdb\_v1\dbfootnote). ChildCIdb is an on-going CCI database collected in collaboration with the school GSD Las Suertes in Madrid (Spain). This database is planned to be extended yearly to enable longitudinal studies. The first version contains children interaction using both finger and stylus. In total, it comprises 438 children aged from 18 months to 8 years, grouped in 7 different educational levels according to the Spanish education system. In particular, 6 different tests are considered in ChildCIdb, grouped in 2 main blocks: \textit{i)} touch, and \textit{ii)} stylus. All tests were designed considering many of the cognitive and neuromuscular aspects highlighted in the state of the art, e.g., the evolution of children's gestures with age. In addition, tests were discussed and approved by neurologists, child psychologists, and educators of the GSD school. The database also considers other children's interesting information such as grades at school, previous experience using mobile devices, attention deficit/hyperactivity disorder (ADHD), prematurity (under 37 weeks gestation), and birthday, among other things. In the present study we focus on the analysis of the time series generated in Test 6 (Drawing Test) of ChildCIdb, in which children have to colour a tree using a pen stylus as good as they can and a maximum time of 2 minutes. We select this test as children have a high degree of freedom to show their motor and cognitive development skills.

In this study, we consider an extension of the current ChildCIdb\_v1 database, including two more acquisition sessions in time. The first one was captured in May 2021 and 376 children participated in the acquisition, adding new young children (18 months). In addition, children who participated in the first version of the database were also acquired again but this time they belong to the next educational level as one year passed since the first acquisition. Children who moved to a higher educational level than the last considered in ChildCIdb (over 8 years old) were excluded from the acquisition. The second acquisition was captured in October 2021 and involved 315 children (similar trend as described before). This extension of ChildCIdb comprises in total 502 different children's metadata and, in total, 1,129 samples of children interaction with mobile devices. Table~\ref{tab:acquisitions} describes the total number of samples collected for each age group. Finally, for completeness, a set of 70 adults aged 25-65 years with fully developed motor and cognitive skills was also captured. This set is not considered in the present work but could be used in future research.

%% file: body/6_proposed_method.tex
\section{Age Group Detection: Proposed Approach}\label{proposed_method}

This section describes our proposed approach for the task of children age group detection based on the automatic analysis of the time series generated while children colouring a tree, i.e., the Drawing Test of the ChildCIdb database. Fig.~\ref{fig:architecture} provides a graphical representation of the proposed approach. We describe next the details of each module. Sec~\ref{time_series_extraction} shows the time series extracted from the Drawing Test. Sec~\ref{classification_metrics} provides the main details of the classification algorithms and metrics studied. Finally, Sec~\ref{time_series_selection} describes the time series selection techniques considered in this study.

% Tabla con series temporales consideradas
\begin{table}[t]
    \caption{Set of time series considered in this work.}
    \label{tab:local_features}
    \resizebox{\linewidth}{!}{%
        \begin{tabular}{c|l}
        \textbf{\#} & \multicolumn{1}{c}{\textbf{Time Serie}} \\ \hline \hline
        1 & X-coordinate: $x_{n}$ \\ \hline
        2 & Y-coordinate: $y_{n}$ \\ \hline
        3 & Stylus pressure: $z_{n}$ \\ \hline
        4 & Path-tangent angle: $\theta_{n}$ \\ \hline
        5 & Path-velocity magnitude: $v_{n}$ \\ \hline
        6 & Log curvature radius: $\rho_{n}$ \\ \hline
        7 & Total acceleration magnitude: $a_{n}$ \\ \hline
        8-14 & First-order derivate of features 1-7: $x'_{n}$, $y'_{n}$, $z'_{n}$, $\theta'_{n}$, $v'_{n}$, $'\rho'_{n}$, $a''_{n}$ \\ \hline
        15-16 & Second-order derivate of features 1-2: $x''_{n}$, $y''_{n}$ \\ \hline
        17 & \begin{tabular}[c]{@{}l@{}}Ratio of the minimum over the maximum speed over a \\ 5-samples windows: $v^{r}_{n}$\end{tabular} \\ \hline
        18-19 & \begin{tabular}[c]{@{}l@{}}Angle of consecutive samples and first-order difference:\\ $\alpha_{n}$, $\alpha'_{n}$\end{tabular} \\ \hline
        20 & Sine: $\sin_{n}$ \\ \hline
        21 & Cosine: $\cos_{n}$ \\ \hline
        22 & Stroke length to width ratio over a 5-samples window: $r^{5}_{n}$ \\ \hline
        23 & Stroke length to width ratio over a 7-samples window: $r^{7}_{n}$ \\ \hline
        24 & \begin{tabular}[c]{@{}l@{}}Indication of whether the child colours the tree inside \\ or outside the boundaries: \bm{$i$}\end{tabular} \\ \hline
        25 & Timestamp at which each sample is taken: $t$
        \end{tabular}
    }
\end{table}

% Análisis Estadístico - DBA
\begin{figure*}[t]
    \begin{center}
       \includegraphics[width=\linewidth, height=5cm]{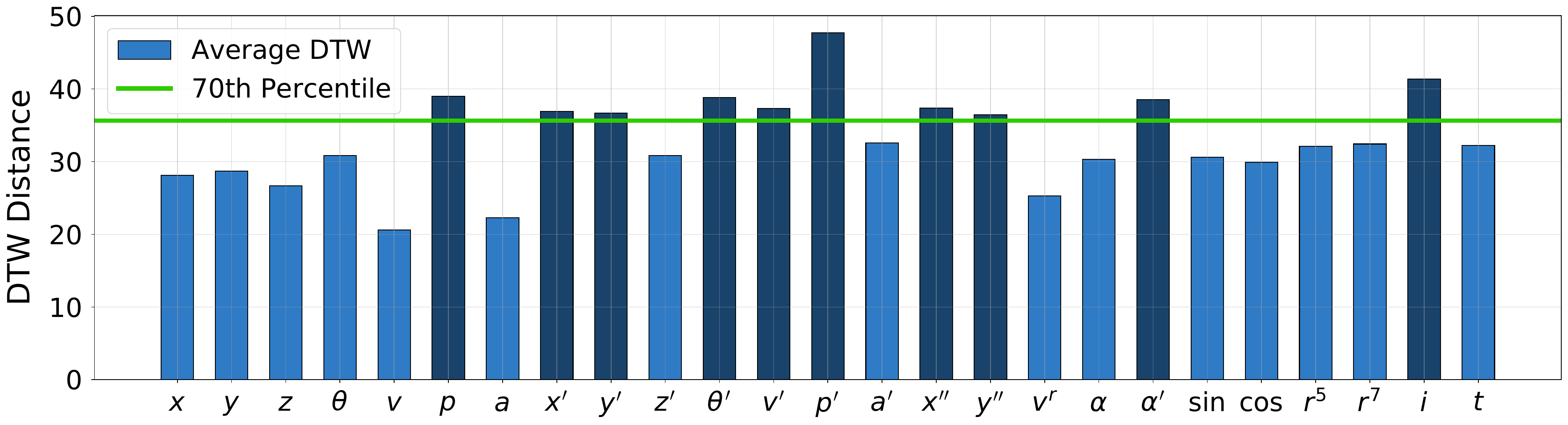}
    \end{center}
    \caption{Statistical Analysis + DBA: Average DTW distance calculated for each time serie considered. The \textbf{\color[HTML]{2fcc00} green line} refers to the 70th percentile of DTW distance. We highlight in \textbf{\color[HTML]{19436b} dark blue} the selected time series.}
    \label{fig:sa_dba}
\end{figure*}

\subsection{Time Series Extraction}\label{time_series_extraction}

Each time a child performs the Drawing Test, a set of 6 time series is captured by the tablet device containing the following information: $x$ and $y$ spatial coordinates, whether the child is colouring inside the tree or not, the pressure performed on the screen using the stylus, timestamp, and type of action performed (pen-down or pen-up). Taking this as a starting point, we extract a set of 25 total time series related to velocity, acceleration, pressure, and stylus position, among others. In particular, 21 time series are based on previous studies on handwriting and signature biometric recognition~\cite{Tolosana2015a}, and the remaining 4 presented in this work are related to stylus pressure and position information. Table~\ref{tab:local_features} shows the final set of time series considered in this study.

\subsection{Classifiers and Metrics}\label{classification_metrics}

\subsubsection{Metrics} we consider 2 metrics to measure the results achieved in the task of children age group detection: 

\begin{itemize}
    \item \textit{Accuracy (\%)}: this metric measures the overall accuracy of the classification system by taking into account only whether the predicted age group of a child is the same as the real one (label).
    
    \item \textit{Age Group Distance (AGD)}: this metric measures the absolute distance between the real age group of a child (label) and the predicted one.
\end{itemize}

\subsubsection{Classification Algorithms}\label{classification} in total, 2 different classifiers are studied as part of our experimental framework. The optimal configuration parameters for each of them depend on the time series selection technique used (Sec.~\ref{time_series_selection}).

% Análisis Estadístico - HMM 2
\setcounter{figure}{3}
\begin{figure*}[!b]
    \begin{center}
       \includegraphics[width=\linewidth, height=5cm]{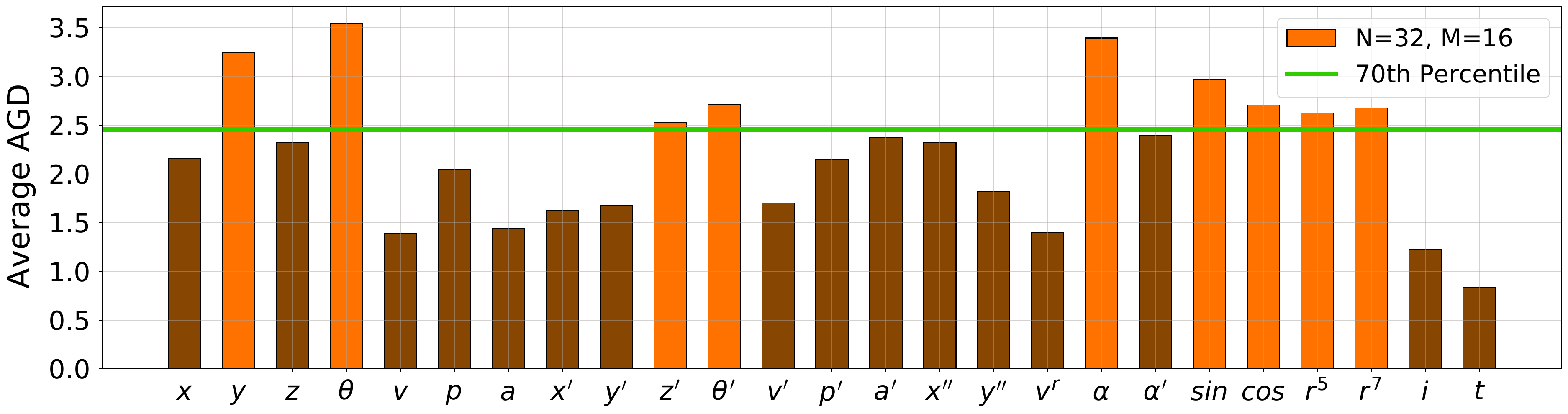}
    \end{center}
    \caption{Statistical Analysis + HMM: Average AGD achieved for each time serie considered. The \textbf{\color[HTML]{2fcc00} green line} refers to the 70th percentile of average AGD. We highlight in \textbf{\color[HTML]{874601} brown} the selected time series.}
    \label{fig:sa_hmm_2}
\end{figure*}

\begin{itemize}
    \item \textit{DTW Barycenter Averaging (DBA)}: this algorithm generates for each time serie and age group the template/prototype that best represent that time serie in that specific age group. As a result, we have 175 prototypes in total (one for each age group, 7, and time series type, 25). DBA computes each prototype by minimising the Dynamic Time Warping (DTW) distance between the extracted time series of the same type and all children of the same age group. Classification is performed by comparing the extracted time series of an input test with each of the generated prototypes, giving as output the age group associated with the smallest DTW distance. The specific implementation considered in this study was presented by Petitjean \textit{et al.} in~\cite{Petitjean2011}.
    
    \item \textit{Hidden Markov Model (HMM)}: this algorithm represents a dual stochastic process, governed by an underlying Markov chain with a finite number of states (\textit{N}) and a random set of symbol-generating functions (\textit{M}), each associated with one state. In contrast to DBA, which is an elastic distance algorithm, HMM is a statistical algorithm. In particular, an HMM model is generated for each age group. HMM models have been widely used in the field of speech recognition and signature verification using time series with accurate results~\cite{Fahad2021, Tolosana2015b, Tolosana2019}. The classification is done by obtaining the probability that the child behaves like most children in that age group. In other words, each child is assigned the age group associated with the most similar HMM model. In our experiments, the configuration parameters \textit{N} and \textit{M} of the HMM models depend on the time series selection technique used. The specific implementation considered in this study is publicly available in HMM-Learn toolbox~\footnote{\url{https://hmmlearn.readthedocs.io/en/stable/}}. 
\end{itemize}

\subsection{Time Series Selection Techniques}\label{time_series_selection}

The following time series selection techniques are studied to choose the most discriminative time series from the 25 total time series considered in the present study:

\begin{itemize}
    \item \textit{Statistical Analysis}: a different statistical analysis of the time series has been conducted according to the classification algorithm used. Focusing on DBA, for each prototype extracted for each time serie, we first calculate the DTW distance with respect to the rest of prototypes of the same time serie but from different age groups in order to see the discriminative power of that time serie with respect to all the age groups (inter-class variability). A similar analysis is carried out for HMM models but using AGD.
    
    \item \textit{Sequential Forward Search (SFS)}: this is a widely used feature selection algorithm that automatically selects an optimal feature subset (time series subset in this case) from the original set using a specific optimisation criteria (AGD in this study). In particular, SFS offers a suboptimal solution because it doesn't take into account all possible combinations, but it does consider correlations between features. The specific implementation considered in this study is publicly available in Scikit-Learn~\footnote{\url{https://scikit-learn.org/stable/}}.

\end{itemize}

% \begin{figure*}[!t]
%     % \vspace{0.5cm}
%     \centering
%     % Análisis Estadístico - HMM 1
%     \begin{minipage}[t]{0.48\linewidth}   
%         \includegraphics[width=\textwidth]{images/sa_hmm_1.pdf}
        
%         \caption{Statistical Analysis + HMM: Results achieved using different HMM parameters. We highlight in \textbf{\color[HTML]{874601} brown} the configuration with the best performance.}
%         \label{fig:sa_hmm_1}
%     \end{minipage}
%     \hfill
%     % SFS (DBA y HMM)
%     \begin{minipage}[t]{0.48\linewidth}   
%         \includegraphics[width=\textwidth]{images/sfs.pdf}
%         \caption{SFS Algorithm for DBA and HMM: Average AGD achieved for DBA and HMM during the execution of SFS in the development stage (training + validation).}
%         \label{fig:sfs}
%     \end{minipage}
% \end{figure*}

% Análisis Estadístico - HMM 1
\setcounter{figure}{2}
\begin{figure}[!b]
    \begin{center}
       \includegraphics[width=\linewidth]{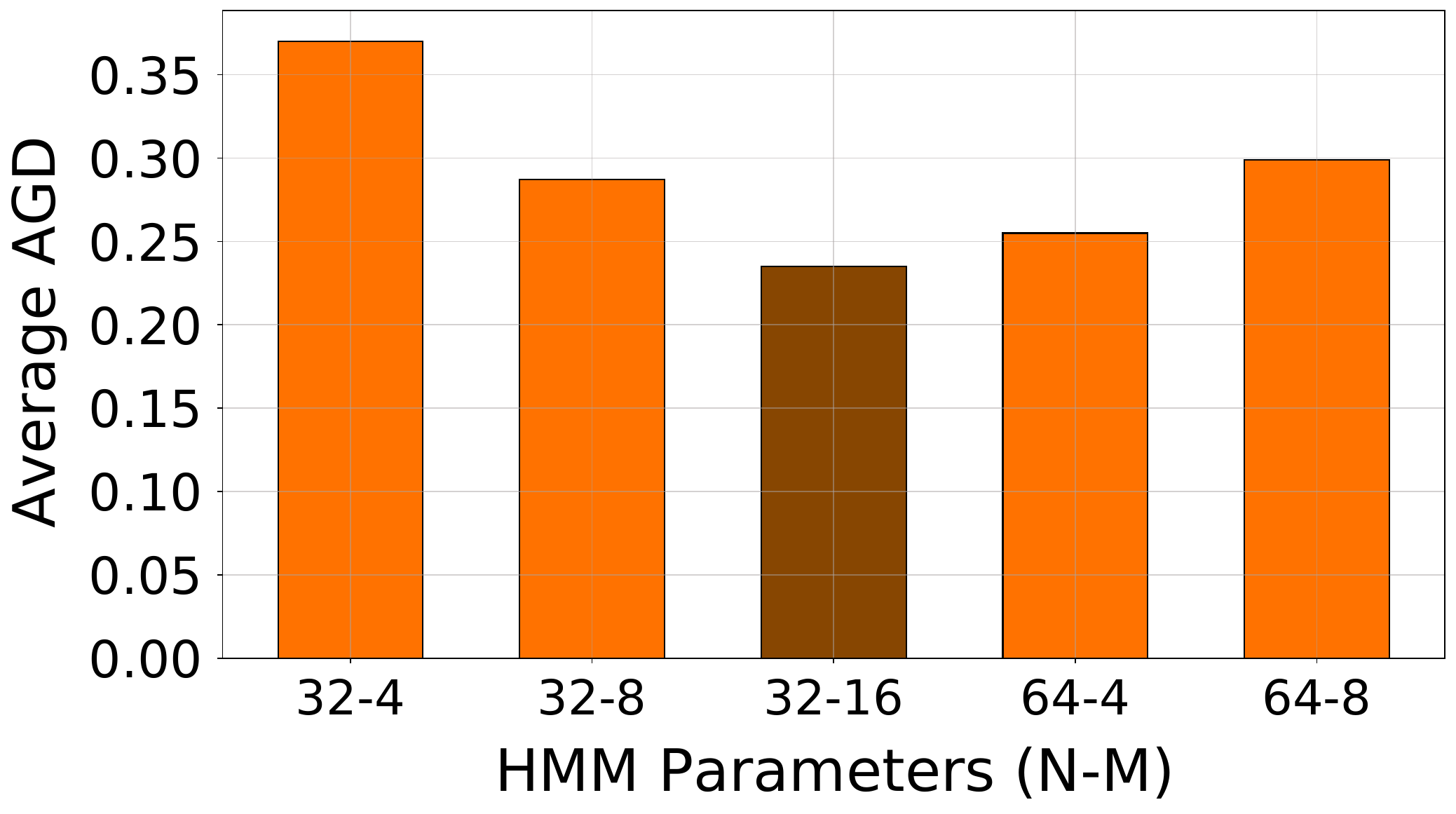}
    \end{center}
    \caption{Statistical Analysis + HMM: Results achieved using different HMM parameters. We highlight in \textbf{\color[HTML]{874601} brown} the configuration with the best performance.}
    \label{fig:sa_hmm_1}
\end{figure}

%% file: body/7_experiments_results.tex
\section{Experiments and results}\label{experiments}

\subsection{Experimental Protocol}

The experimental protocol considered has been designed to detect the age group of the children based on their educational level through the analysis of their interactions with mobile devices. In particular, 7 age groups are considered according to the Spanish education system as we can see in Table~\ref{tab:age_range_group}. ChildCIdb is divided into development (80\% of the children) and evaluation (20\% remaining children) datasets, where children in each age group are equally distributed in number and gender to allow an objective comparison of the results obtained. In addition, the development dataset is divided in 2 different subsets: training (80\%) and validation (20\%) datasets. During the development stage, the children considered in the evaluation dataset are excluded. Only the training subset is used to train the different age group detection systems, while the validation subset is used to test their performance. Finally, the final evaluation dataset is used to test the age group detection systems under realistic conditions, considering children unseen during the development stage. All experiments are run on a machine with an Intel i7-9700 processor and 32GB of RAM.

\subsection{Experimental Results}

\subsubsection{\textbf{Time Series Selection}} this section analyses which are the most discriminative and robust time series for each classification algorithm considered, using both SFS and statistical analysis selection techniques. For this purpose, we only consider the training and validation datasets. The following 4 cases are studied:

\begin{itemize}
    \item \textit{Statistical Analysis for DBA}: we create 7 prototypes (one for each age group) for each time serie using DBA. Then, for each time serie, we calculate the DTW distance between the 7 age groups using the corresponding prototypes (inter-class variability). The higher the average DTW distance is, the higher the discriminative power of that time serie will be to detect different age groups. We select those time series whose average DTW distance is above the 70th percentile (see Fig.~\ref{fig:sa_dba}).
    
    \item \textit{Statistical Analysis for HMM}: first, we create a single HMM model and consider different HMM parameter configurations using all the time series. The HMM parameters with the best performance (lowest average AGD) in the validation dataset are \textit{N}=32 and \textit{M}=16 (see Fig.\ref{fig:sa_hmm_1}). Then, we generate a new HMM model for each time serie, and finally, we calculate the average AGD for the task of children age group detection. The lower the average AGD is, the higher the discriminative power of that time serie will be. We select those time series whose average AGD is in the 70th percentile (see Fig.~\ref{fig:sa_hmm_2}).
    
    \item \textit{SFS Algorithm for DBA}: first, SFS selects the time serie with the highest discriminative power (lowest AGD). Then, SFS selects the time series that obtains the lowest AGD by performing combinations of 2. This procedure is performed successively until the result on the validation dataset gets worse. Fig.~\ref{fig:sfs} provides a graphical representation of the development stage. The best performance of the system is achieved with 8 time series ($v^r$, $i$, $\theta$, $x'$, $y''$, $\theta'$, $a$, $\alpha'$).
    
    \item \textit{SFS Algorithm for HMM}: the same procedure described before is applied to HMM. The HMM parameters with the best performance in the validation dataset are \textit{N}=64 and \textit{M}=8. The best performance of the system (Fig.~\ref{fig:sfs}) is achieved with 9 time series ($t$, $r^7$, $z$, $x''$, $z'$, $x$, $y$, $\theta'$, $p'$). 
\end{itemize} 

\subsubsection{\textbf{Results}} this section analyses the performance achieved on the final evaluation dataset for the task of children's age group detection, considering the optimal configurations described in the previous section. Table~\ref{tab:results} shows the results achieved in terms of accuracy (\%) and average AGD (distance measured in age groups). For completeness, we show in Fig.~\ref{fig:mage_results_by_group} the average AGD obtained for each age group and classification algorithm considered.

Focusing first on DBA, it seems that using prototypes for representing each age group of children is not an accurate approach. On the one hand, the time series selection technique used (i.e., Statistical Analysis or SFS) does not greatly affect the results obtained. In terms of accuracy (\%), the best result achieved is 33.19\%. On the other hand, focusing on the average AGD, although the DBA system is not able to detect correctly the age group, the distance between the label and the predicted class is not so bad (average AGD of 0.99 age groups for the SFS approach, i.e., average distance lower than one group). This aspect can be better observed in Fig.~\ref{fig:mage_results_by_group}. It seems that the higher average AGD values happen with the intermediate groups (4 to 6). We believe this can be produced due to at these educational levels children face a developmental stage full of multiple educational changes and learning phases. In particular, children in group 3 have just moved from nursery school to pre-school and those in group 6 are about to move from pre-school to primary education. This generates challenges in classification.

% SFS (DBA y HMM)
\setcounter{figure}{4}
\begin{figure}[t]
    \begin{center}
       \includegraphics[width=\linewidth]{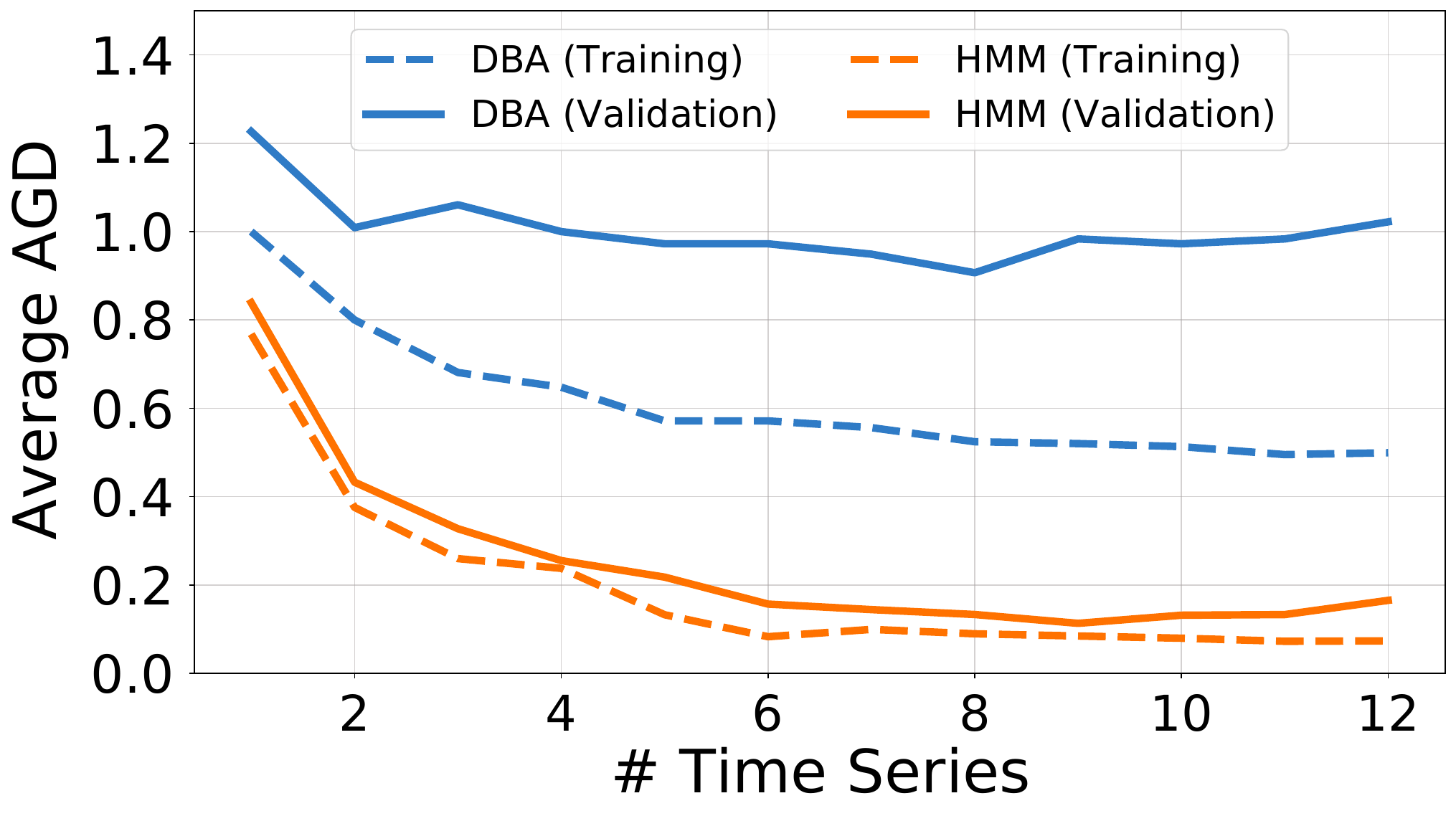}
    \end{center}
    \caption{SFS Algorithm for DBA and HMM: Average AGD achieved for DBA and HMM during the execution of SFS in the development stage (training + validation).}
    \label{fig:sfs}
\end{figure}

Analysing the HMM system, we can see in Table~\ref{tab:results} that different results are achieved depending on the time series selection technique. The SFS approach outperforms the statistical analysis in terms of accuracy (85.39\% vs. 75.22\%) and average AGD (0.17 vs. 0.31). Again, as can be seen in Fig.~\ref{fig:mage_results_by_group}, the intermediate groups are also the most challenging ones for the HMM models. Nevertheless, HMM models obtain a very low average AGD, with less than 0.2 age groups. Hence, HMM seems to be an accurate statistical approach to model the children patterns in the time domain by exploiting the correlation of the chosen time series.

\subsubsection{\textbf{Comparison with the State-of-the-Art}}

Finally, we compare the results obtained with the state of the art. It is important to remark that, as indicated in Table~\ref{tab:comparison} of the article, a direct comparison with previous studies in the literature is not feasible as different tasks, databases, and experimental protocols are usually considered. Nevertheless, and although the tasks considered in previous studies are simpler (e.g., detecting children from adults), we include them for a better understanding of the results achieved with our proposed approach. For example, in~\cite{VeraRodriguez2020} the authors obtained an accuracy of 93\% on the task of detecting children from adults using an automatic detection system based on finger touch interaction. Similar results were achieved by Vatavu \textit{et al.} in~\cite{Vatavu2015}, over 80\%, using a feature set extracted from the interaction of adults and children on mobile devices. 

It is also interesting to compare the results achieved in the present work with our previous approach~\cite{Tolosana2022a} focusing on the same Drawing Test, although a simpler task was considered. Concretely, we considered 3 age groups (i.e., 1-3 years, 3-6 years, and 6-8 years) instead of the 7 age groups considered in the present article (i.e., one for each educational level as indicated in Table~\ref{tab:age_range_group}). In~\cite{Tolosana2022a}, we presented an approach based on 148 global features. We achieved results above 90\% accuracy on the children age group detection task. This approach based on global features presented a less discriminative way of detecting the age group associated with a child than the approach proposed in the present article based on the analysis of the complete realization process of the Drawing Test through the time signals. The present approach (local features or time series) captures more detailed patterns of children’s interaction, allowing us to consider a more challenging scenario in which to detect each educational level of children (7 possible groups). For completeness, we include in Table~\ref{tab:local_vs_global} a comparison between the results obtained using the proposed approach (local features) and the previous one presented in~\cite{Tolosana2022a} (global features) focusing on the same task, i.e., detection of the 7 possible age groups of children. To obtain the results using the global features approach, we consider the best configuration obtained in~\cite{Tolosana2022a}. Therefore, comparing the results obtained in the present work (accuracy results over 85\% and average AGD of 0.17 age groups) with the results achieved by using the global features approach (30.09\% accuracy and average AGD of 1.62 age groups), we can conclude that the automatic analysis of the time series (local features) generated while colouring the tree provides much more discriminative information of the children age group compared with approach based on the global features considered in~\cite{Tolosana2022a}.

% Tabla con resultados finales en terminos de accuracy y AGD
\begin{table}[t]
\caption{Results achieved in terms of accuracy (\%) and average AGD over the final evaluation set for the Drawing Test of ChildCIdb. We highlight in \textbf{bold} the configuration that provides the best results.}
\label{tab:results}
\begin{tabular}{c|cc|cc}
 & \multicolumn{2}{c|}{\textbf{Statistical Analysis}} & \multicolumn{2}{c}{\textbf{SFS}} \\[+3pt] \cline{2-5} 
 & \textbf{DBA}\rule{0pt}{12pt} & \textbf{HMM} & \textbf{DBA} & \textbf{HMM} \\ \hline
\textbf{Accuracy (\%)}\rule{0pt}{12pt} & 33.19 & 75.22 & 32.74 & \textbf{85.39} \\
\textbf{Average AGD}\rule{0pt}{12pt} & 1.05 & 0.31 & 0.99 & \textbf{0.17}
\end{tabular}
\end{table}

% Tabla con resultados finales en terminos de accuracy y AGD
\begin{table}[t]
\caption{Results of the global and local features approaches in terms of accuracy (\%) and average AGD over the final evaluation set for the Drawing Test of ChildCIdb. We highlight in \textbf{bold} the approach that provides the best results.}
\label{tab:local_vs_global}
\begin{tabular}{c|c|c}
\textbf{} & \textbf{Accuracy (\%)} & \textbf{Average AGD} \\ \hline
\textbf{\begin{tabular}[c]{@{}c@{}}Global Features\\ \cite{Tolosana2022a}\end{tabular}}\rule{0pt}{20pt}  & 30.09 & 1.62 \\ [+10pt] \hline
\textbf{\begin{tabular}[c]{@{}c@{}}Local Features\\ (Proposed Approach)\end{tabular}}\rule{0pt}{20pt} & \textbf{85.39} & \textbf{0.17}
\end{tabular}
\end{table}

Finally, the results achieved in this study prove the high potential of combining CCI and automatic analysis of time series for the task of children age group detection. This can benefit many children-related applications, for example, towards an age-appropriate environment with the technology.

%% file: body/8_conclusion_feature_works.tex
\section{Conclusion and Future Work}\label{conclusions}

This study proposes a novel Children-Computer Interaction (CCI) approach for the task of children age group detection in order to benefit many children-related applications, for example, towards an age-appropriate environment with the technology. In particular, we have focused on the automatic analysis of the time series generated from the interaction of children with mobile devices. We have focused on a specific test of ChildCIdb, Drawing Test, where children have to colour a tree over a pen stylus tablet. After that, we extract 25 total time series related to spatial, pressure, and kinematic information.

Our proposed approach has been studied using two different time series selection techniques to choose the most discriminative ones: \textit{i)} a statistical analysis, and \textit{ii)} an automatic feature selection algorithm called Sequential Forward Search (SFS). In addition, two classification algorithms have been considered for the task of children age group detection: \textit{i)} DTW Barycenter Averaging (DBA), and \textit{ii)} Hidden Markov Models (HMM).

% Resultados finales con DBA y HMM usando Análisis Estadístico y SFS
\begin{figure}[t]
    \begin{center}
       \includegraphics[width=\linewidth]{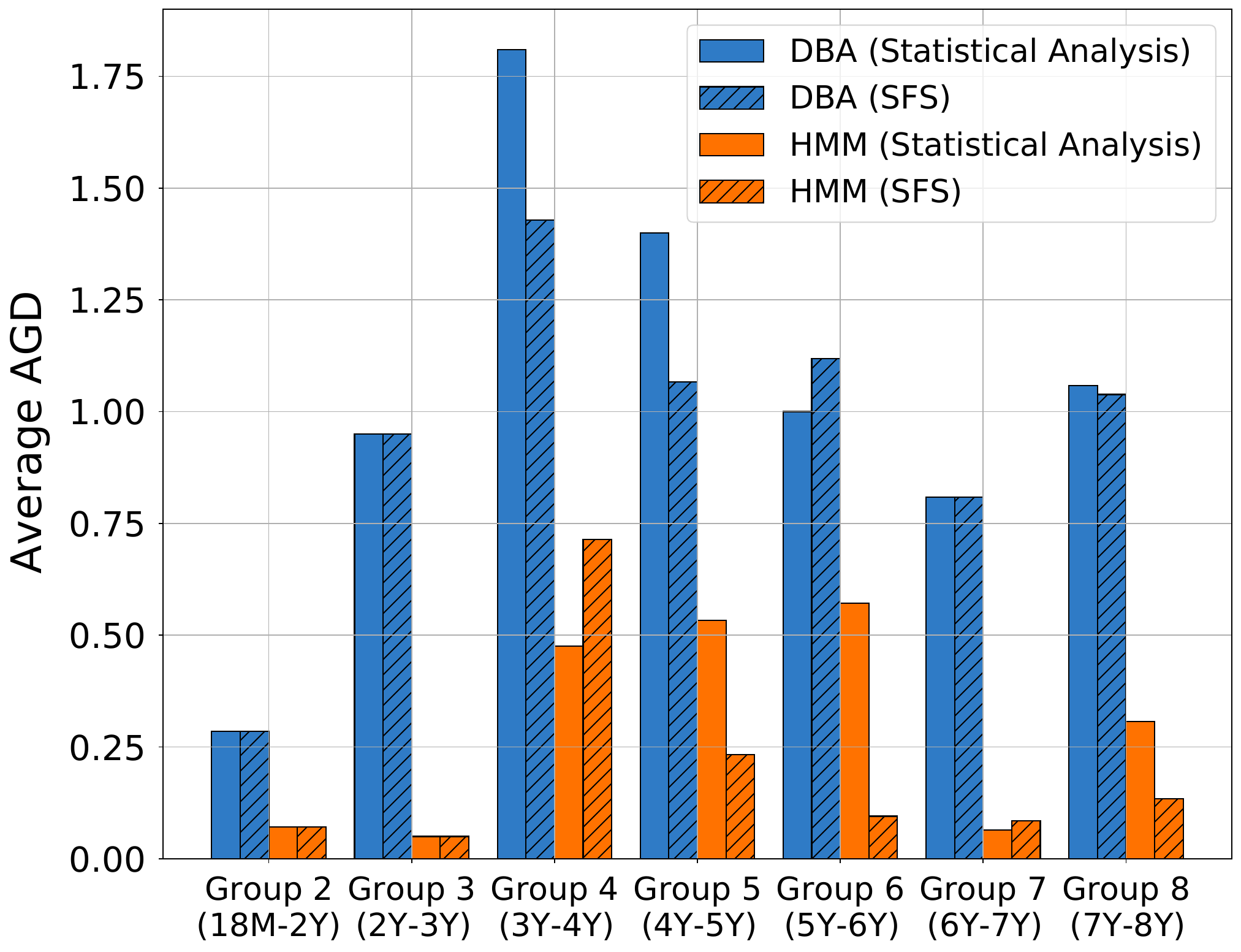}
    \end{center}
    \caption{Average AGD achieved for each of the age groups, classification algorithms, and time series selection techniques in the final evaluation dataset of ChildCIdb.}
    \label{fig:mage_results_by_group}
\end{figure}

Analysing the results obtained, our proposed HMM + SFS approach has achieved better accuracy results for the task of children age group detection (7 possible groups) compared to previous studies in the literature, which focused on global features by considering simpler tasks such as detecting children from adults or group children in only 3 age groups. In particular, the proposed approach achieves an average AGD lower than 0.2 age groups and accuracy over 85\%. These results prove the high potential of combining an innovative approach based on CCI and automatic analysis of time series (local features) for the task of children age group detection.

Future works will be oriented to: \textit{i)} a longitudinal analysis of the children, studying the motor and cognitive evolution of them through the interaction with the tests of ChildCIdb, \textit{ii)} consider the ChildCIdb database in other areas of e-Health and e-Learning, for example in terms of privacy~\cite{Delgado-Santos2022, Melzi2022}, and \textit{iii)} analyse the relationship of children's metadata (grades, ADHD, prematurity, etc.) to their interaction with mobile devices.

%% file: body/9_acknowledgements.tex
\section*{Acknowledgements}

This work has been supported by project INTER-ACTION (PID2021- 126521OB-I00 MICINN/FEDER) and HumanCAIC (TED2021-131787B-I00 MICINN). J.C. Ruiz-Garcia is supported by the Madrid Government (Comunidad de Madrid-Spain) under the Multiannual Agreement with Autonomous University of Madrid in the line Encouragement of the Research of Young Researchers, in the context of the V PRICIT (Regional Programme of Research and Technological Innovation). This is an on-going project carried out with the collaboration of the school GSD Las Suertes in Madrid, Spain.

%% file: body/10_biography.tex
\vspace{-10mm}

\begin{IEEEbiography}[{\includegraphics[width=1in,height=1.25in,clip,keepaspectratio]{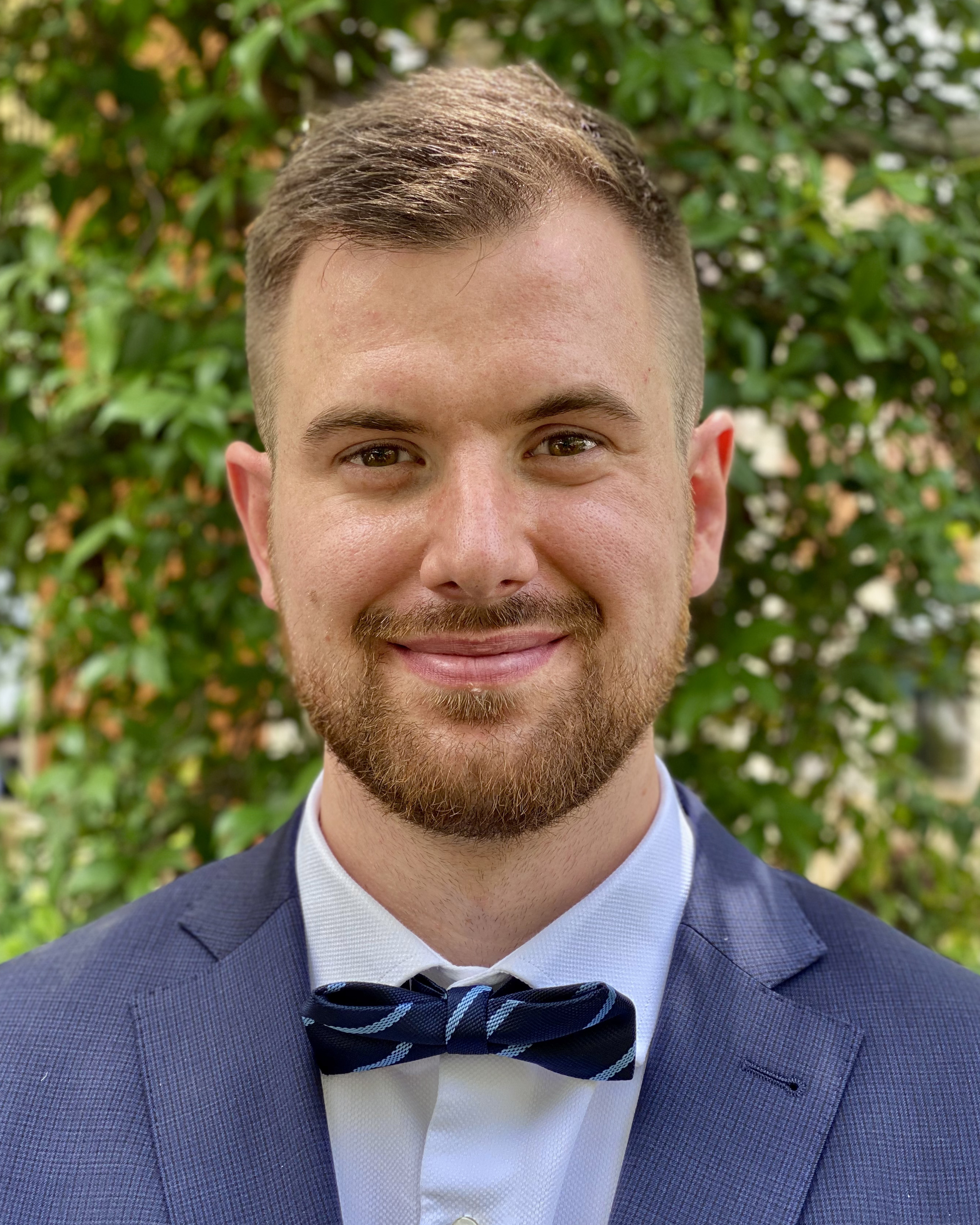}}]{Juan Carlos Ruiz-Garcia} received his B.Sc. degree in Computer Science Engineering in 2019 from the Universidad de Granada and got the M.Sc. degree in Research and Innovation in 2021 with the award of excellence from the Universidad Autonoma de Madrid, where he is currently pursuing a PhD degree in Computer and Telecommunication Engineering. In addition, in April 2020, he joined the Biometrics and Data Pattern Analytics - BiDA Lab as Pre-Doctoral Researcher at the same university. His research interests are mainly focused on the use of machine learning for e-Learning, e-Health, Human-Computer Interaction (HCI), and automatic Fall Detection Systems (FDS).
\end{IEEEbiography}

\vspace{-2mm}

\begin{IEEEbiography}[{\includegraphics[width=1in,height=1.25in,clip,keepaspectratio]{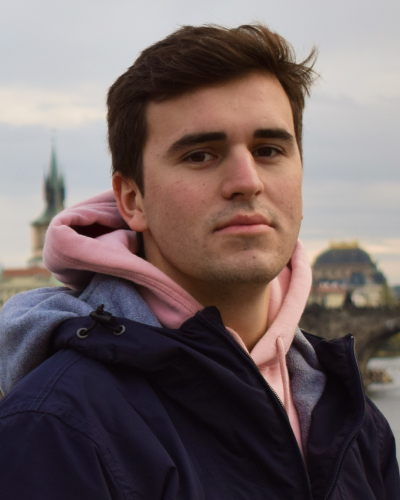}}]%
{Carlos Hojas}
received his joint Bachelor Degree in Mathematics and Computer Science in 2022 from the Universidad Autónoma de Madrid, Spain. He did his Computer Science Thesis within the Biometrics and Data Pattern Analytics - BiDA Lab, focusing on time series classification with machine learning, receiving an honorable mention in the best final project contest of Deloitte.
\end{IEEEbiography}

\begin{IEEEbiography}[{\includegraphics[width=1in,height=1.25in,clip,keepaspectratio]{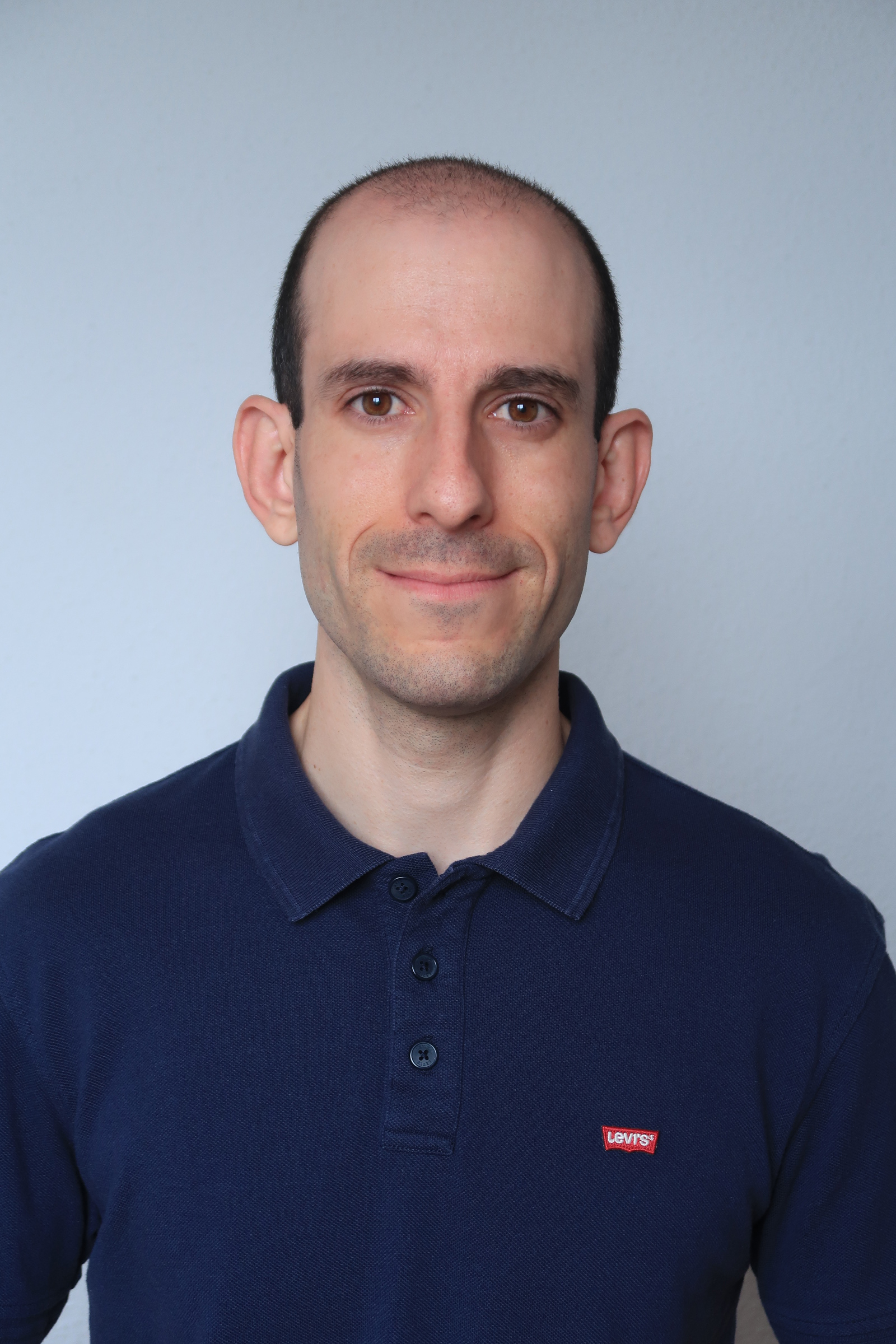}}]%
{Ruben Tolosana} received the M.Sc. degree in Telecommunication Engineering, and the Ph.D. degree in Computer and Telecommunication Engineering, from Universidad Autonoma de Madrid, in 2014 and 2019, respectively. In 2014, he joined the Biometrics and Data Pattern Analytics – BiDA Lab at the Universidad Autonoma de Madrid, where he is currently Assistant Professor. He is a member of the ELLIS Society (European Laboratory for Learning and Intelligent Systems), Technical Area Committee of EURASIP (European Association For Signal Processing), and Editorial Board of the IEEE Biometrics Council Newsletter. His research interests are mainly focused on signal and image processing, Pattern Recognition, and Machine Learning, particularly in the areas of DeepFakes, Human-Computer Interaction, Biometrics, and Health. Dr. Tolosana is actively involved in several National and European projects focused on these topics. He has also organized several workshops and challenges in top conferences such as WAMWB (MobileHCI 2023), KVC (BigData 2023), MobileB2C (IJCB 2022) and SVC-onGoing (ICDAR 2021). Dr. Tolosana has also received several awards such as the European Biometrics Industry Award (2018) from the European Association for Biometrics (EAB) and the Best Ph.D. Thesis Award in 2019-2022 from the Spanish Association for Pattern Recognition and Image Analysis (AERFAI).
\end{IEEEbiography}

\begin{IEEEbiography}[{\includegraphics[width=1in,height=1.25in,clip,keepaspectratio]{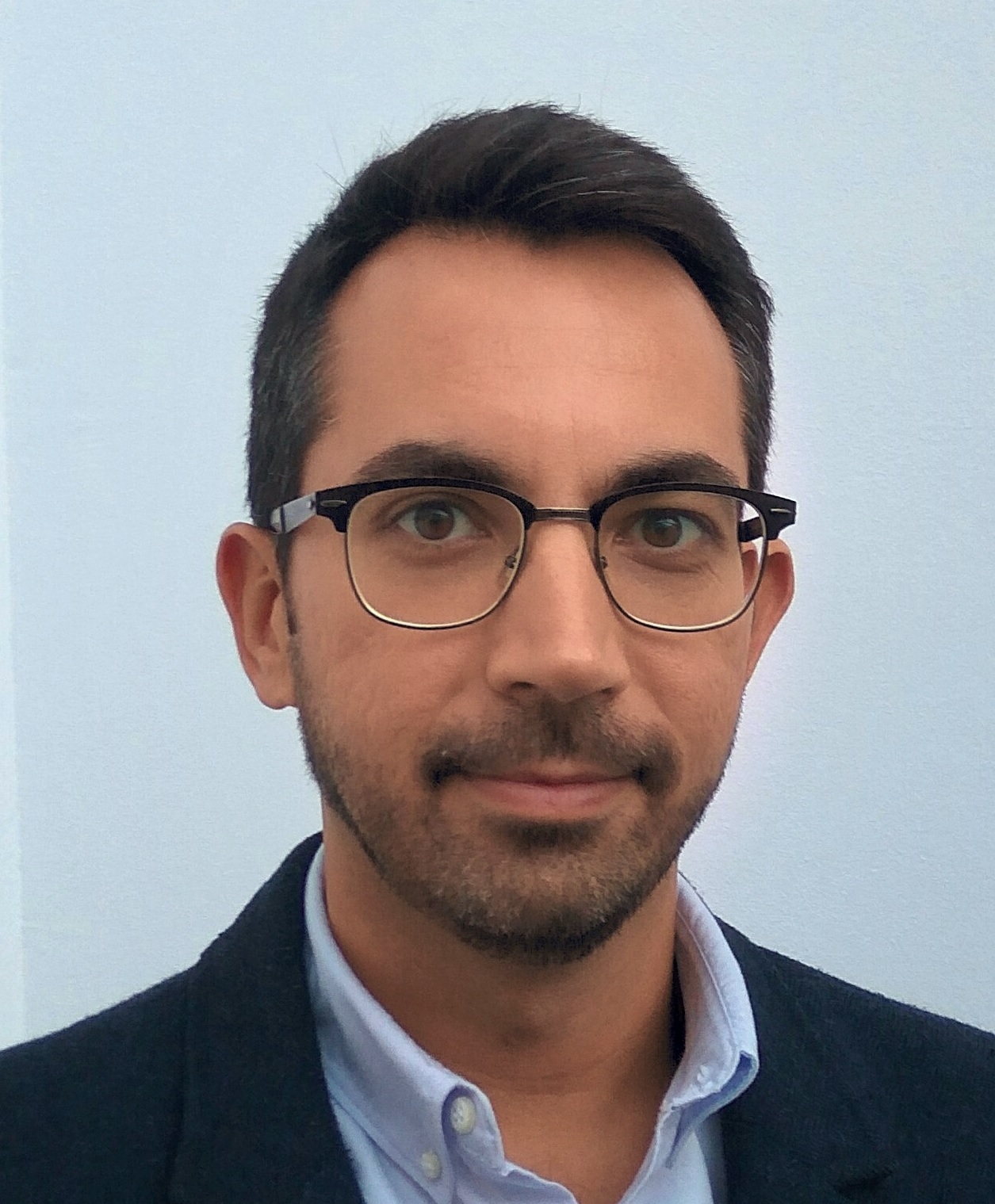}}]{Ruben Vera-Rodriguez}received the M.Sc. degree in telecommunications engineering from Universidad de Sevilla, Spain, in 2006, and the Ph.D. degree in electrical and electronic engineering from Swansea University, U.K., in 2010. Since 2010, he has been affiliated with the Biometric Recognition Group, Universidad Autonoma de Madrid, Spain, where he is currently an Associate Professor since 2018. His research interests include signal and image processing, AI fundamentals and applications, HCI, forensics, and biometrics for security and human behavior analysis. Dr. Vera-Rodriguez is actively involved in several National and European projects focused on these topics. He is author of more than 150 scientific articles published in international journals and conferences. He has served as Program Chair for some international conferences such as: IEEE ICCST 2017, CIARP 2018, ICBEA 2019 and AVSS 2022. He has also organized several workshops and challenges in top conferences such as WAMWB (MobileHCI 2023), KVC (BigData 2023), MobileB2C (IJCB 2022) and SVC-onGoing (ICDAR 2021). Ruben has received a Medal in the Young Researcher Awards 2022 by the Spanish Royal Academy of Engineering among other awards, and he is member of ELLIS Society since 2023.
\end{IEEEbiography}

\begin{IEEEbiography}[{\includegraphics[width=1in,height=1.25in,clip,keepaspectratio]{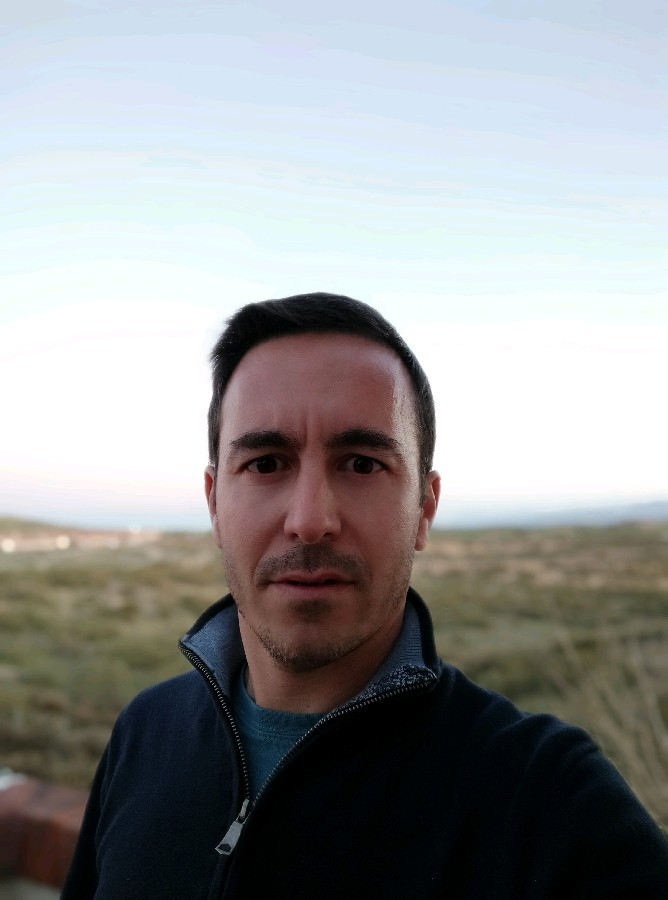}}]{Aythami Morales} received his M.Sc. degree in Electrical Engineering in 2006 from Universidad de Las Palmas de Gran Canaria. He received his Ph.D. degree in Artificial Intelligence from La Universidad de Las Palmas de Gran Canaria in 2011. He performs his research works in the BiDA Lab – Biometric and Data Pattern Analytics Laboratory at Universidad Autónoma de Madrid, where he is currently an Associate Professor (CAM Lecturer Excellence Program). He is member of the ELLIS Society (European Laboratory for Learning and Intelligent Systems). He has performed research stays at the Biometric Research Laboratory at Michigan State University, the Biometric Research Center at Hong Kong Polytechnic University, the Biometric System Laboratory at University of Bologna and Schepens Eye Research Institute (Harvard Medical School). He is author of more than 100 scientific articles published in international journals and conferences, and 2 patents. A. Morales is supported by the Madrid Government (Comunidad de Madrid-Spain) under the Multiannual Agreement with Universidad Autónoma de Madrid in the line of Excellence for the University Teaching Staff in the context of the V PRICIT (Regional Programme of Research and Technological Innovation).
\end{IEEEbiography}

\begin{IEEEbiography}[{\includegraphics[width=1in,height=1.25in,clip,keepaspectratio]{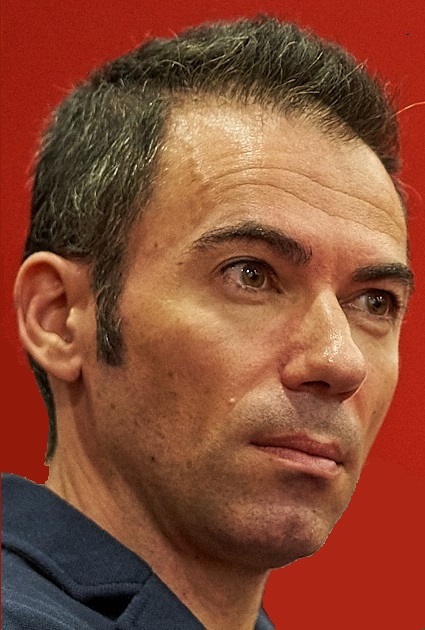}}]{Julian Fierrez} received the MSc and the PhD degrees from Universidad Politecnica de Madrid, Spain, in 2001 and 2006, respectively. Since 2004 he is at Universidad Autonoma de Madrid, where he is Associate Professor since 2010. His research is on signal and image processing, AI fundamentals and applications, HCI, forensics, and biometrics for security and human behavior analysis. He is Associate Editor for Information Fusion, IEEE Trans. on Information Forensics and Security, and IEEE Trans. on Image Processing. He has received best papers awards at AVBPA, ICB, IJCB, ICPR, ICPRS, and Pattern Recognition Letters; and several research distinctions, including: EBF European Biometric Industry Award 2006, EURASIP Best PhD Award 2012, Miguel Catalan Award to the Best Researcher under 40 in the Community of Madrid in the general area of Science and Technology, and the IAPR Young Biometrics Investigator Award 2017. Since 2020 he is member of the ELLIS Society.
\end{IEEEbiography}

\begin{IEEEbiography}[{\includegraphics[width=1in,height=1.25in,clip,keepaspectratio]{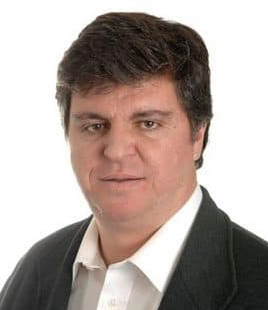}}]{Javier Ortega-Garcia} received the M.Sc. degree in electrical engineering and the Ph.D. (cum laude) degree in electrical engineering both from Universidad Politecnica de Madrid, Spain, in 1989 and 1996, respectively. He is the co-Director of BiDALab (Biometric \& Data Pattern Analytics Lab) and a Full Professor at the Signal Processing Chair, within the Escuela Politecnica Superior, Universidad Autonoma de Madrid, where he teaches bio-signal processing, biometric recognition and human-device interaction courses. From 2018 he is Fellow of IEEE “for contributions to biometrics for forensic speaker verification”, and of IAPR (International Association for Pattern Recognition). He has authored over 300 peer-reviewed international high-impact contributions, including book chapters, journal articles, and conference papers. His research interests are focused on biometric authentication, human-device interaction, bio-signal classification and pattern matching for AI-based applications in security, healthcare, wellness and banking domains, among others. He was appointed as chair of the “2004 Odyssey: The Speaker Recognition Workshop”, the 3rd “IAPR International Conference on Biometrics” (ICB) in 2009, and the 6th IAPR ICB in 2013; the 2016 IEEE ICCSI Carnahan Conference on Security Technologies; and IEEE AVSS-2022, the 18th IEEE Int. Conf. on Advanced Video and Signal-Based Surveillance in Madrid Nov-Dec 2022. 
\end{IEEEbiography}

\vspace{-115mm}

\begin{IEEEbiography}[{\includegraphics[width=1in,height=1.25in,clip,keepaspectratio]{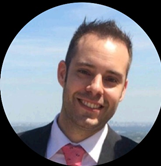}}]{Jaime Herreros-Rodriguez (JHR)} received the degree in Medicine in 2006 from Universidad Autónoma de Madrid, the tittle of neurologist in 2010 and he was awarded the title of Doctor in Medicine from the Universidad Complutense de Madrid (2019) with a distinction Cum Laude given unanimously for his doctoral thesis on migraine. He is also author of several publications in migraine and parkinsonism. He has collaborated with different research projects related to many neurological disorders, mainly Alzheimer and Parkinson's disease. JHR is a neurology and neurosurgery proffesor in CTO group, since 2008.
\end{IEEEbiography}